\documentclass[conference]{IEEEtran}
\IEEEoverridecommandlockouts
\usepackage{cite}
\usepackage{amsmath,amssymb,amsfonts}
\usepackage{algorithmic}
\usepackage{svg}
\usepackage[hyphens]{url}
\usepackage{hyperref}
\usepackage{graphicx}
\usepackage{textcomp}
\usepackage{xcolor}
\usepackage{paralist}
\usepackage{fancyhdr}
\usepackage{lipsum}% generate text for the example
\usepackage{makecell}
\usepackage{comment}

\def\BibTeX{{\rm B\kern-.05em{\sc i\kern-.025em b}\kern-.08em
    T\kern-.1667em\lower.7ex\hbox{E}\kern-.125emX}}
    
\fancypagestyle{firstpagefooter}{%
  \fancyhf{}
  
}

\makeatletter % changes the catcode of @ to 11
\newcommand{\linebreakand}{%
  \end{@IEEEauthorhalign}
  \hfill\mbox{}\par
  \mbox{}\hfill\begin{@IEEEauthorhalign}
}
\makeatother % changes the catcode of @ back to 12

\begin{document}
\title{Tapping into Privacy: A Study of User Preferences and Concerns on Trigger-Action Platforms}

\author{\IEEEauthorblockN{Piero Romare}
\IEEEauthorblockA{\textit{Department of Computer Science and Engineering} \\
\textit{Chalmers University of Technology}\\
Gothenburg, Sweden \\
pieror@chalmers.se}
\and
\IEEEauthorblockN{Victor Morel}
\IEEEauthorblockA{\textit{Department of Computer Science and Engineering} \\
\textit{Chalmers University of Technology}\\
Gothenburg, Sweden \\
morelv@chalmers.se}
\and
\linebreakand
\IEEEauthorblockN{Farzaneh Karegar}
\IEEEauthorblockA{\textit{Department of Mathematics and Computer Science} \\
\textit{Karlstad University} \\
Karlstad, Sweden \\
farzaneh.karegar@kau.se}
\and
\IEEEauthorblockN{Simone Fischer-Hübner}
\IEEEauthorblockA{\textit{Department of Computer Science and Engineering} \\
\textit{Chalmers University of Technology \& Karlstad University}\\
Gothenburg \& Karlstad, Sweden \\
simonefi@chalmers.se, simofihu@kau.se}
}

%\author{\IEEEauthorblockN{1\textsuperscript{st} Piero Romare}
%\IEEEauthorblockA{\textit{Department of Computer Science and Engineering} \\
%\textit{Chalmers University of Technology}\\
%Gothenburg, Sweden \\
%pieror@chalmers.se}
%\and
%\IEEEauthorblockN{2\textsuperscript{nd} Victor Morel}
%\IEEEauthorblockA{\textit{Department of Computer Science and Engineering} \\
%\textit{Chalmers University of Technology}\\
%Gothenburg, Sweden \\
%morelv@chalmers.se}
%\and
%\IEEEauthorblockN{3\textsuperscript{rd} Farzaneh Karegar}
%\IEEEauthorblockA{\textit{Department of Computer Science} \\
%\textit{Karlstad University}\\
%Karlstad, Sweden \\
%farzaneh.karegar@kau.se}
%\and
%\IEEEauthorblockN{4\textsuperscript{th} Simone Fischer-Hübner}
%\IEEEauthorblockA{\textit{Department of Computer Science} \\
%\textit{Karlstad University}\\
%Karlstad, Sweden \\
%simone.fischer-huebner@kau.se}
%}

\maketitle

\begin{abstract}
The Internet of Things (IoT) devices are rapidly increasing in popularity, with more individuals using Internet-connected devices that continuously monitor their activities.
This work explores privacy concerns and expectations of end-users related to Trigger-Action platforms (TAPs) in the context of the Internet of Things (IoT). TAPs allow users to customize their smart environments by creating rules that trigger actions based on specific events or conditions. As personal data flows between different entities, there is a potential for privacy concerns. In this study, we aimed to identify the privacy factors that impact users' concerns and preferences for using IoT TAPs. To address this research objective, we conducted three focus groups with 15 participants and we extracted nine themes related to privacy factors using thematic analysis. Our participants particularly prefer to have control and transparency over the automation and are concerned about unexpected data inferences, risks and unforeseen consequences for themselves and for bystanders that are caused by the automation.
The identified privacy factors can help researchers derive  
predefined and selectable profiles of privacy permission settings for IoT TAPs that represent the privacy preferences of different types of users as a basis for designing usable privacy controls for IoT TAPs. 
\end{abstract}


\thispagestyle{firstpagefooter}

\section{Introduction}\label{sec:introduction}
In recent years, end users are becoming increasingly involved as high-level programmers to connect and personalize technological tools.  
The usage of Trigger-Action platforms (TAPs), which allow users to customize their own smart environments, is becoming more popular. On these TAPs, Trigger-Action applications let users set the rules that indicate when specific activities will be performed. These rules follow an if-else logic -- also called Event-Condition-Action (ECA)~\cite{demeure_2017_eca, desolda_2017_filter}, an illustration of which can be found in the TAP IFTTT, which stands for ``If This Then That''. The applications can be employed in the Internet of Things (IoT) between at least two connected entities that can be devices, e.g. smart bulbs and/or services such as Dropbox. The connection between the two entities is mediated through TAPs (e,g., IFTTT, Microsoft Power Automate, Zapier) hosting the applications.
These platforms, because they offer user-friendly interfaces building elaborate workflows which can speed up end-user digital tasks, have therefore gained recent traction from academia.
% It can be illustrated briefly by the following: ``\textit{if} a new email with a meeting, \textit{then} create an event in the calendar''. 

As a matter of fact, since users' personal data flows between different entities that trigger events or perform actions, these functionalities may introduce privacy concerns and expectations. Also, the privacy preferences arising from these Trigger-Action applications, where one entity automatically triggers actions involving personal data to be performed on another entity,  may differ from regular IoT environments, in which traditionally only one entity was acting as a provider of a service to which users subscribe and may therefore put trust in.
% to which the users subscribe. 
%Specific attitudes and requirements from the users would imply different needs that play a 
%critical role, but which can provide if addressed more effective use of personal data in the context of Trigger-Action apps.
Eliciting privacy factors, including privacy preferences and concerns, which matter for users will reveal privacy needs that play a role for different types of users. If these factors are well addressed, it may lead to a privacy-preserving, trustworthy, and thus more effective use of personal data by Trigger-Action applications.

%Piero, I gave it a try to rewrite the following part. It should be connected to the sentence you wrote above. I will keep it as a comment so you can see if you want to use it:

%Therefore, in this paper we aim to address the following research question:

% \textit{RQ1: What are the privacy factors that play a role in users' concerns and preferences for using IoT TAPs?}

%To address our research question, we did exploratory work in the form of three focus groups with 15 participants in total to extract privacy factors. In the focus groups, we discussed participants' opinions about IoT TAPs in general including the factors they consider important when using such platforms and the perceived benefits and problems and also their opinions related to specific IFTTT scenarios. Analysing the qualitative data collected using thematic analysis led to the extraction of privacy factors as nine themes that give insights into what may affect users' privacy decisions when using such platforms and their privacy concerns and perceptions. 

% \newline
We present our exploratory work to investigate and compare the privacy expectations and concerns of IoT Trigger-Action applications to address new hypotheses according to the opinions of the participants with respect to the general IoT context. We focus on users' views by stimulating their evaluation between risks and benefits. 
\newline We defined the following Research Question (RQ): \begin{itemize}
    \item \textit{What are the privacy factors that play a role in users' concerns and preferences for using IoT TAPs?}
\end{itemize}

Identified privacy factors will serve as input for the design of a quantitative follow-up user segmentation study for deriving typical privacy personas for IoT TAPs. The derived privacy personas will then provide input for defining profiles of privacy permission settings for IoT TAPs that represent the privacy preferences of different types of users. Privacy profiles meeting users' personas can be offered to them for easily selecting or adapting their permission settings.

For our study reported in this paper, we were particularly interested in investigating how far privacy factors for IoT TAPs that matter for users  go beyond and complement privacy factors for ordinary IoT environments that were analyzed and previously elicited in the literature.

Three focus group discussions on privacy concerns and expectations concerning TAPs were conducted involving 15 participants. A facilitator presided over the group, the discussions are audio-recorded and transcriptions are made for evaluation.
%From our qualitative data, we show the insights, suggestions revealing preferences and and concerns of the participants. Using thematic analysis, we detect privacy factors that matter for individuals for the use of IoT TAPs.
Analyzing the qualitative data collected using thematic analysis led to the extraction of privacy factors as nine themes that matter for individuals when using IoT TAPs.
In particular, our participants prefer to have control over data before the final action occurs without having a fully automated procedure, they consider the transparency over automation and data recipients and protection for users and bystanders, and they care about the potential misuse and unexpected consequences.

\textit{Contribution} Our findings shed additional light on users' expectations of privacy for these growing platforms. We particularly derive users' privacy factors for using IoT TAPs and discuss how far they may differ from those factors that matter for users in general IoT environments. 
%including only one entity as a service provider. 
We discuss how our findings can provide input for designing usable privacy-preserving controls that meet users' privacy preferences for IoT TAPs.

%We may be able to add new design options to the survey factors found in the literature with a future quantitative study. 

\textit{Organization} In Section~\ref{sec:background} we present previous works in more detail. We explain in Section~\ref{sec:methods} the methodology of how we conducted the focus groups and in Section~\ref{sec:results} we show the results in terms of privacy concerns and preferences from the participants. We discuss our results and how it goes beyond related findings and conclude the paper in Section~\ref{sec:discussion} and in Section~\ref{sec:conclusion} respectively.

\section{Background and Related Work}\label{sec:background}
\subsection{IoT apps}
IoT applications (in our context, IoT Trigger-Action applications) are simple pieces of code connecting IoT devices (e.g. home assistants, smart cars) with online services (e.g. Twitter or Google).
Popular platforms for these apps include IFTTT, Zapier, and Microsoft Flow.
A specificity of these IoT apps is that they are programmed by end-users, anyone can submit their own app on the platform.

\begin{figure}[!ht]
    \centering
    \includegraphics[scale=.5]{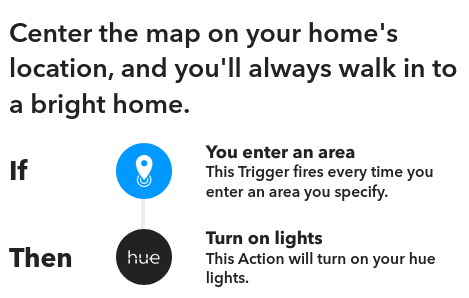}
    \caption{Example of an IoT application on the IFTTT platform.}
    \label{fig:hue}
\end{figure}

An IoT app connects different entities through a \textit{recipe}, which consists of a trigger, an action and eventually filters.
An example of such a recipe for IFTTT (denoted an \textit{applet}) is "Turn your lights on automatically as you arrive home", where the trigger is "You enter an area - This Trigger fires every time you enter an area you specify." and the action "Turn on lights - This Action will turn on your hue lights.", as Figure~\ref{fig:hue} illustrates.~\footnote{\url{https://ifttt.com/applets/HTak4X5f-turn-your-lights-on-automatically-as-you-arrive-home}}
Here, in order for the smart bulb to turn on, the IFTTT platform has to be aware of the user's position, and in some cases, the data can be shared with numerous third parties.

Which data is collected and who has access to it is not always obvious nor minimized~\cite{ahmadpanahlazytap}.
As a matter of fact, authorizing the location for one IoT app on the \textit{IFTTT mobile app} can disclose one's location to other IoT apps (regardless of its utility and lawfulness), and your precise location can be collected even if only your coarse location is necessary.
Therefore, the data flow enabled by these IoT apps can raise serious privacy concerns for users.
This is particularly relevant in the context of the personalization of IoT services and devices, which often involves users as developers.
It is applicable in many environments like the smart home~\cite{ur_practical_2014,salovaara_2021_home}, the smart city~\cite{varun_2022_city,conti_2014_city}, or wearables~\cite{mi_empirical_2017}. 
%For end-users there exist specific programming needs~\cite{brich_2017_needs} and mental models~\cite{huang_supporting_2015} that have been investigated through user studies.

% \subsection{IoT contexts}

\subsection{Related Work: Privacy in the Context of IoT}
\label{Related}
%The effectiveness of TAPs interfaces for end-users simplifies control~\cite{derussis_2015_guidelines,corcella_2019_tool,zhao_understanding_2021} of apps even with tangible interface~\cite{tada_2016_tangible}.
%Previous works investigated the privacy perception in the smart home context~\cite{geeng_control_2019,abbott2022privacyll,zeng_2019_understading} and evaluated the privacy and integrity risk on real users of IFTTT applets~\cite{cobb_how_2020}.
%It takes a lot of effort for the end-users to read privacy policies and their comprehension is quite low~\cite{vu_2007_policies}. On the other hand, regulators refer to them as a contract between companies and users. Through user testing, a uniform consent form with standard layouts across services could educate users to focus on the pertinent details of what data they share~\cite{karegar_2020_notices} because of a decrease of the learnability costs CHECK THIS W FARZANEH. The term privacy notices have been used to refer to situations in which the users of services or connected device know what, how, how long, when, where and why their data and personal information are collected (CITATION HERE IF NEEDED). The challenges of the privacy notices, reported in~\cite{schaub_2015_notices}, are their complexity, lack of choices, fatigue, and decoupling. While privacy notices inform people, privacy choices let users control parts of such users' data. They may be both provided at the same time or suffered from dark patterns~\cite{nouwens_2020_dark}. Similarly,

The recent proliferation of IoT devices has presented end users with a number of privacy challenges. Helping end-users to understand pertinent security and privacy issues in IoT settings is a research direction which still needs to be explored~\cite{paterno_2019_future}. Moreover, users need usable means for managing privacy permissions. In this section, we will summarize previous studies conducted in this area. Our own work, which complements and enhances the existing research will provide important input for designing usable permission systems.

Privacy choices are often hard to understand~\cite{kelly_2012_permissions} and if denied, there can be no access to the service. 
Feng et al. define the features that compose the privacy choices as type, functionality, time, channel and modality~\cite{feng_2021_choices}. From these choices, the users provide permissions to the service or device to use their data. People's data may be collected silently and in previously unanticipated circumstances~\cite{lopez_2017_privacy}.
Furthermore, the complexity of the IoT ecosystem increases as the quantity of available IoT devices~\cite{barricelli_2015_grow} and tools for the End-User Development (EUD)~\cite{fogli_2016_eudtools} grow. Indeed, owners of IoT devices care substantially less about their data because consumers neglected security due to their unfamiliarity and complexity~\cite{william_2017_boring}. The potential privacy repercussions and ownership control of IoT data are mostly unexplored~\cite{weber_2015_iot}.
An analysis ran on a dataset of over 19000 IFTTT existing recipes on secrecy and integrity violations categorized around 50\% of these recipes as potentially unsafe~\cite{surbatovich_2017_analysis}. Policy breaches, unintended flows access control, revocation and privilege are frequent privacy violations in TAPs generated apps~\cite{mahadewa_2021_weaknesses}. A survey regarding IFTTT smart home applications has found how contextual factors influence and play a role in end users' concerns~\cite{saeidi_if_2022}.  A focus group has been conducted to examine the factors impacting individual IoT use concerns and how those aspects impact the dynamics of privacy management~\cite{padyab_2018_fg}. 
With qualitative and quantitative data on IoT privacy concerns in IoT~\cite{emami-naeini_privacy_nodate}, the findings show the users preference of having control over their data, that should be anonymized, and with short retention times. The link between the perceived level of trust as well as the benefits, and the perceived privacy risk with the information sensitivity in the IoT services may increase the users' data disclosure~\cite{kim_2019_trust}. The trustworthy factors that the users perceived relating to smart system providers include privacy concerns like transparency, unauthorized data access, user control, oversight from government and independent institutions, trust in the manufacturer~\cite{kulyk_2020_trust}. In the smart home context, users were interviewed about their perceptions of smart home privacy risks, and measures adopted to protect their privacy. They suggest that who should have access to their smart home data is based on the perceived benefits of giving such access, but they are unaware of the machine learning inference that might expose sensitive information~\cite{zheng_2018_iot}. 

In summary, privacy concerns that can be commonly derived from the related work above include: 
\begin{inparaenum}[i)]
    \item unauthorized data sharing, unauthorized data access and leakage~\cite{william_2017_boring, kulyk_2020_trust, saeidi_if_2022},
    \item undesired consequences (e.g., surveillance, data selling, discrimination, safety risks)~\cite{william_2017_boring, naeini_2023_city, saeidi_if_2022},
    \item profiling/aggregation and sensitive data inferences~\cite{padyab_2018_fg, saeidi_if_2022},
    \item out of context data inferences and secondary use~\cite{padyab_2018_fg},
    \item vulnerabilities, 
    \item data breaches~\cite{padyab_2018_fg, emami-naeini_privacy_nodate, naeini_2023_city}, and
    \item perceived harm and risks~\cite{kim_2019_trust,naeini_2023_city}.
\end{inparaenum}

Privacy preferences that were mostly in common by the related work include:
\begin{inparaenum}[i)]
    \item transparency and notifications~\cite{kulyk_2020_trust, emami-naeini_privacy_nodate, kim_2019_trust, naeini_2023_city}, 
    \item user control~\cite{kulyk_2020_trust, emami-naeini_privacy_nodate}, 
    \item data minimization through anonymous data collection or short data retention ~\cite{padyab_2018_fg, emami-naeini_privacy_nodate, naeini_2023_city},
    \item bystanders privacy~\cite{marky2020visitors,cobb2021incidental,yao_2019_bystanders},
    \item place of storage~\cite{padyab_2018_fg},
    \item reputation and trust in device and manufacturer~\cite{kulyk_2020_trust, padyab_2018_fg, emami-naeini_privacy_nodate, kim_2019_trust, zheng_2018_iot, naeini_2023_city}, 
    \item purposes and perceived benefits vs. risks~\cite{kim_2019_trust, zheng_2018_iot, naeini_2023_city}, 
    \item data security and technical privacy protection~\cite{zheng_2018_iot, naeini_2023_city}, and
    \item privacy / safety tradeoffs~\cite{naeini_2023_city}.
\end{inparaenum}

To the best of our knowledge, our work is the first that used qualitative research methods in the form of focus groups to explore user's privacy preferences and concerns for IoT TAPs in more detail to gain in-depth insights into the privacy factors that matter for users and for what reasons. It thus goes beyond the previous related work that either only considers regular IoT environments or uses only quantitative methods for determining privacy factors related to IoT TAPs (such as ~\cite{saeidi_if_2022}) without exploring in depth the motivations why such factors matter. We will discuss in section~\ref{sec:discussion} how our findings on privacy factors for IoT TAPs are complementing the related work.

\section{Methods}
\label{sec:methods}
A focus group is a group interview organized by a facilitator who moderates a discussion on a specific topic. 
An assistant moderator is also present during the session to help the main facilitator in taking notes. 
The debate is left to the participants once the facilitator introduces the subject. 
It usually counts five to seven participants and takes 1-2 hours. We follow the guidelines provided in~\cite{krueger_2002_fginterviews} with a procedure close to~\cite{valdez_2018_perspective}. 

For this study, we conducted three Focus Groups (FG) with 15 participants in total (FG1: 5, FG2: 6, FG3: 4) on-site in two different universities (Chalmers University of Technology and Karlstad University), lasting for two hours each, from December to March 2023.
We describe our methodology in this section, including how the focus groups were conducted and how the data was extracted and annotated. We recorded the audio of the focus groups using Zoom. Through Zoom, automatic transcription is one feature that allows the conversion from audio to text. We cross-checked the automatic transcriptions with recorded audio files to check their correctness.  

\subsection{Recruitment}
We recruited 15 participants in total for our three focus groups by advertising through the university campus, university employees not working in the privacy or security area, and personal networks. 
The message for the recruitment was a brief description of the TAPs platforms with a high-level explanation of the planned data processing for our analysis in which we advertised our study as an exploration of users' opinions regarding and perceptions of IoT application scenarios involving IoT Trigger-Action Platforms.
We did not mention any terminology concerning privacy concerns and risks to avoid a priming effect. 
In each focus group, we asked participants to fill in the demographics data: age group, gender, the highest level of education achieved, and their subjective opinion about their IT knowledge on a scale from 1 to 5 (poor, fair, good, very good, excellent). Table~\ref{tab4} shows the details of our participants' demographics. 

\begin{table}
\caption{Participants Demographics}\label{tab4}
\begin{tabular}{|l|l|l|l|l|l|}
\hline
FG & Participant & Age & Gender & Education & IT Knowledge \\
\hline
1 & P1 & 25-34 & Male & Master & Very Good\\ \hline
1 & P2 & 25-34 & Female & Master & Fair \\ \hline
1 & P3 & 25-34 & Male & Master & Very Good\\ \hline
1 & P4 & 25-34 & Female & Master & Fair \\ \hline
1 & P5 & 25-34 & Male & Master & Good \\ \hline
2 & P6 & 18-24 & Female & Bachelor & Poor\\ \hline
2 & P7 & 18-24 & Female & Bachelor & Very Good\\ \hline
2 & P8 & 25-34 & Female & Bachelor & Good\\ \hline
2 & P9 & 18-24 & Female & Bachelor & Good\\ \hline
2 & P10 & 18-24 & Female & Bachelor & Good\\ \hline
2 & P11 & 18-24 & Female & Bachelor & Fair\\ \hline
3 & P12 & 25-34 & Male & PhD & Excellent\\ \hline
3 & P13 & 45-54 & Male & PhD & Excellent\\ \hline
3 & P14 & 55+ & Female & High School & Poor \\ \hline
3 & P15 & 45-54 & Male & Bachelor & Very Good \\
\hline
\end{tabular}
\end{table}
 
\subsection{Legal and ethical considerations}
We asked for and obtained an informed consent from the participants before each session. All collected data is securely pseudonymized and protected following the rules of the General Data Protection Regulation (GDPR).
We received confirmation from the university Data Protection Officer (DPO) about the cloud recording through Zoom, in particular, the servers that stored the recording are placed in Europe, Germany and Sweden.
The study design was reviewed and accepted by Karlstad University's ethics advisor.
The participants were each compensated on the same day as the focus group with a coupon for the university store or canteen. 

\subsection{Course of the focus groups}\label{FGsteps}
Briefly put, each focus group session includes
\begin{inparaenum}[a)]
\item a prologue session in which we welcomed participants, asked demographic questions and request their consent,
\item an introductory session for describing the context (IoT TAP) and persona to be used during the focus groups,
\item a session for a general discussion related to IoT TAPs,
\item a session on discussion related to specific TAP scenarios, and 
\item a sorting task to conclude the focus group.
\end{inparaenum}

We organized a break of 15 minutes with beverages and snacks between parts c and d. In the following of this section, we provide more details concerning steps \textit{c} to \textit{e}. Appendix~\ref{appendix-1} presents the questions and the follow-up questions we asked our participants in different parts of our focus groups.

\subsubsection{Description of the persona}
Before starting the discussions -- general and scenarios-based -- we introduced the persona. 
The persona is an artificial user that has the double goal of preserving the participant's personal experience and facilitating the context of TAPs in everyday users' life. 
We called it ``Alex'', a gender-neutral name, who is described with needs and usage of IFTTT apps: 
\begin{itemize}
    \item Alex would like to automatize their life and digital life tasks using IoT applications;
    \item Alex has already automatized and connected some of their IoT devices and services;
    \item Alex is thinking that it is harder and harder to manually keep track of what is happening on all of these services and devices, what data they receive, what they do and how one can affect another;
    \item Alex will help us to understand needs, experiences, behaviours, and goals regarding IoT applications.
\end{itemize} 

\subsubsection{Introduction to the topic and general discussion}
We expanded the context of our discussions, in particular by presenting what is the Internet of Things (IoT) and TAPs in more depth. 
After some real cases examples of usage, we assessed the participants' understanding by asking them to provide other realistic examples (see Appendix~\ref{appendix-1} for the question asked in this regard).
%and see Table~\ref{tab3} for participants' provided examples). 
We decided to allow the discussion for 45 minutes after which a break of 15 minutes was made. 
Through all the discussions, the participants were authorized and encouraged to discuss both realistic and idealistic scenarios (i.e., a scenario that doesn't exist already). 

\subsubsection{Scenarios}
We converged the discussion by proposing three scenarios of our own for getting a more concrete view from the participants. 
We created these scenarios based on the following rule: they must be described as an \textit{if this then that} recipe involving at least two entities, where one of them is an IoT device; this rule ensured the validity of an IoT app setting including a data flow between different entities, thus hinting possible privacy concerns.

We devised scenarios so as to illustrate the different features of IoT apps. The scenarios are happening in different contexts for TAPs and with the different types of actors and relationships among them. With these hypothetical scenarios, given the limited time in this study, we could still cover the broad spectrum of trigger-action applications that might exist in the real world and their influence on users' preferences.
Mainly, the features (see Table~\ref{tab1}) are implicitly written in the scenarios, but the trigger-action formula is presented together with the figurative representation to the participants of our focus groups. 

We proposed two first scenarios (1.a and 1.b), related to the smart home. Since during the first focus group we noticed fewer participants' involvement in 1.a, we introduced 1.b which was applied in the context of the smart home as well. The second scenario is related to the smart city and the third one is related to wearables.

\begin{table*}[!h]
\centering
\caption{Features in the scenarios.}\label{tab1}
\begin{tabular}{|l|l|l|l|l|l|}
\hline
Feature & Introductory Scenario & Scenario 1a & Scenario 1b & Scenario 2 & Scenario 3 \\
\hline
Type of data & Personal & Personal & Personal & Sensitive & Personal\\
Type of controller & Private & Private & Private & Private & Public\\
Role of the IoT device(s) in the scenario & Trigger & Trigger and Action & Trigger & Trigger & Trigger\\
%Purpose of data collection & Location & Financial & Entertainment & Financial \\
%Retention time & Timeline (HOW TO EXPRESS THIS?) \\
Context & Wearables & Smart Home & Smart Home & Wearables & Smart City \\
%If This Then That & Scenario related formula \\
\hline
\end{tabular}
\end{table*}

\begin{itemize}
    \item Introductory Scenario: 
    Alex is using a smartwatch that keeps track of his/her sleep patterns. If the smartwatch detects that Alex had less than one hour of deep sleep last night, then his/her early meetings will be cancelled on his/her Google calendar (see Figure~\ref{fig1}).
    \begin{verbatim}
        If  sleep < 1h 
        Then delete morning meetings
    \end{verbatim}
        %\begin{figure}[!h]
        %\includegraphics[scale=0.08]{Figures/FigSleep.png}
        %\caption{Introductory Scenario: since Alex did not sleep well the morning meetings are deleted.}
        %\label{fig1}
        %\end{figure}
    %\begin{enumerate}
    %    \item  
    \item Scenario 1a: 
    Alex is moving from home to his/her office. The location is shared from his/her smartphone to his/her smart vacuum cleaner robot. It starts cleaning the living room because it knows that the room is now free and dirty. To help participants better understand each scenario, we showed a related figure on a big screen. Figure~\ref{fig2a} shows what was presented for scenario 1a. The formula in this scenario is 
    \begin{verbatim}
        If location == office 
        Then clean
    \end{verbatim}
        \begin{figure}[!h]
        \includegraphics[scale=0.08]{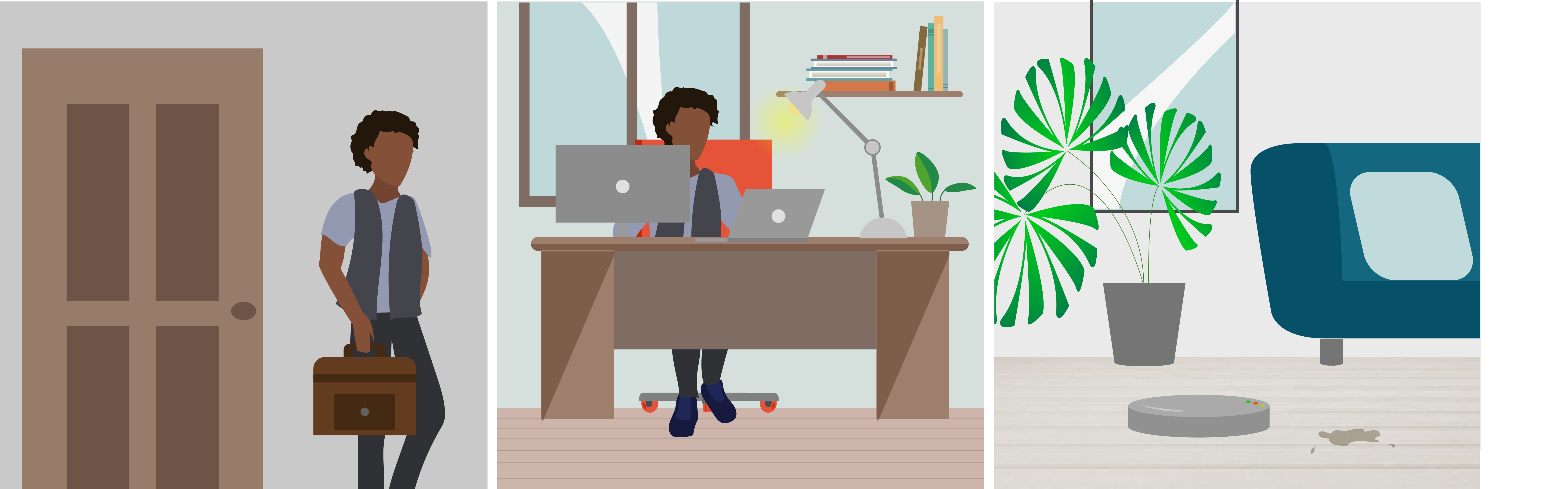}
        \caption{Scenario 1a: Alex is at work while the house is cleaned.} \label{fig2a}
        \end{figure}
    \item Scenario 1b: Alex's smart fridge detects that there is no more orange juice, and it then sends a message to his/her google cloud notes to update his/her shopping list and add orange juice. The orange juice will be deleted from the list if the fridge detects it or after one month (see Figure~\ref{fig2b}). The formula for this scenario is 
    \begin{verbatim}
        If orangejuice == FALSE 
        Then add to shoppinglist
    \end{verbatim}
        \begin{figure}[!h]
        \includegraphics[scale=0.08]{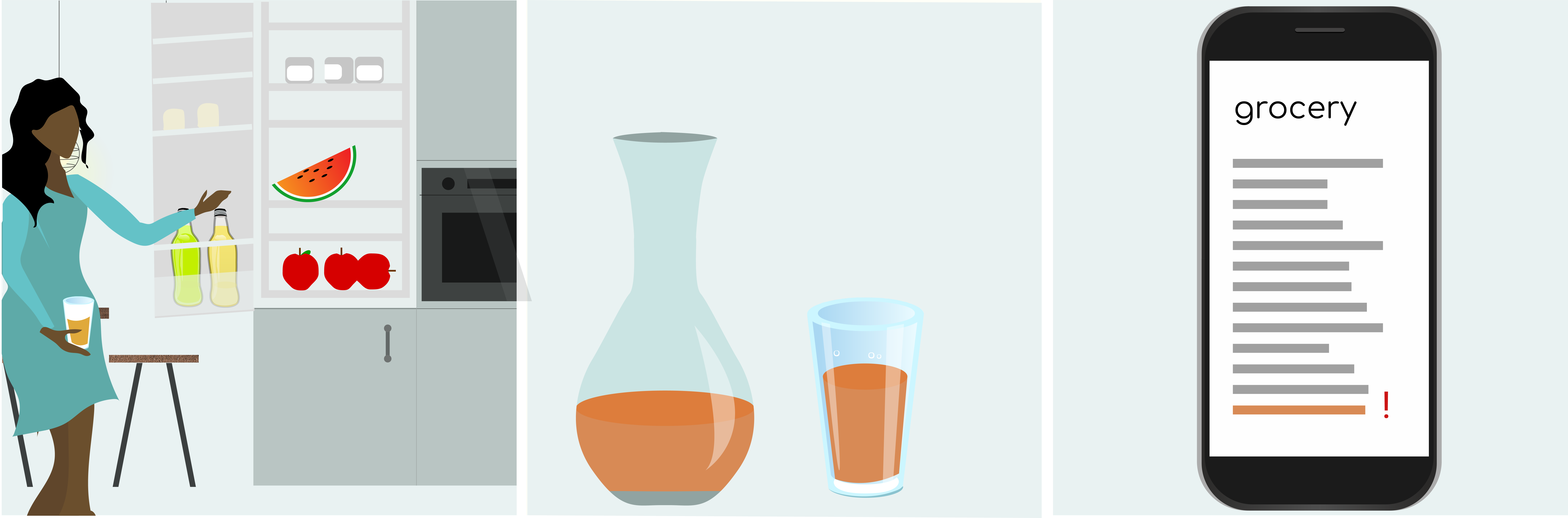}
        \caption{Scenario 1b: Alex is drinking orange juice while the shopping list is updating.} \label{fig2b}
        \end{figure}
    \item Scenario 2: Alex is a person very active on social media. Using his/her smart glasses, Alex records videos that are automatically uploaded to his/her TikTok profile (see Figure~\ref{fig3}). The formula is 
    \begin{verbatim}
        If newvideo == TRUE 
        Then upload on TikTok
    \end{verbatim}
        \begin{figure}[!h]
        \includegraphics[scale=0.08]{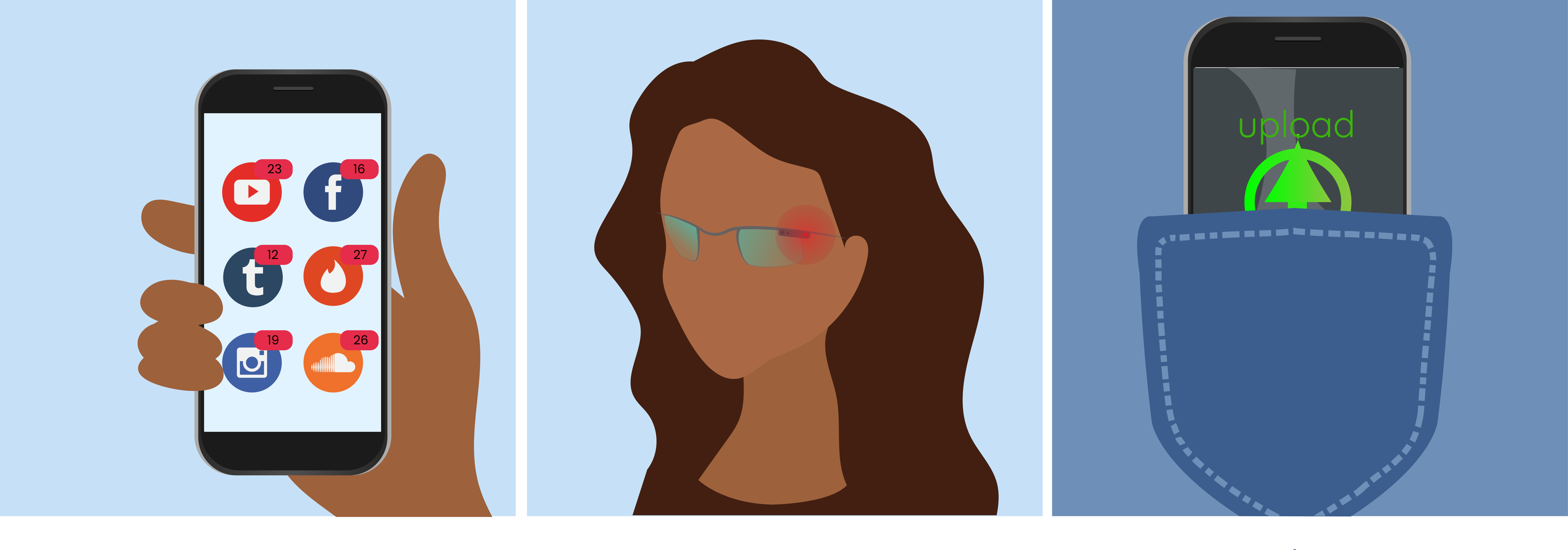}
        \caption{Scenario 2: Alex is uploading smartglasses videos on online social media.} \label{fig3}
        \end{figure}
    \item Scenario 3: Alex's city has implemented a public smart water grid (get a further explanation) that allows public professionals to monitor the quantity and quality of water. Alex's private smart water meter shares the consumption with the smart grids which allow the city to send out itemized bills. After some expensive bills, Alex has set a consumption threshold and if Alex goes above it then the next usage will be with colder water (see Figure~\ref{fig4}).
    \begin{verbatim}
        If consumption > threshold 
        Then decrease temperature
    \end{verbatim}
        \begin{figure}[!h]
        \includegraphics[scale=0.08]{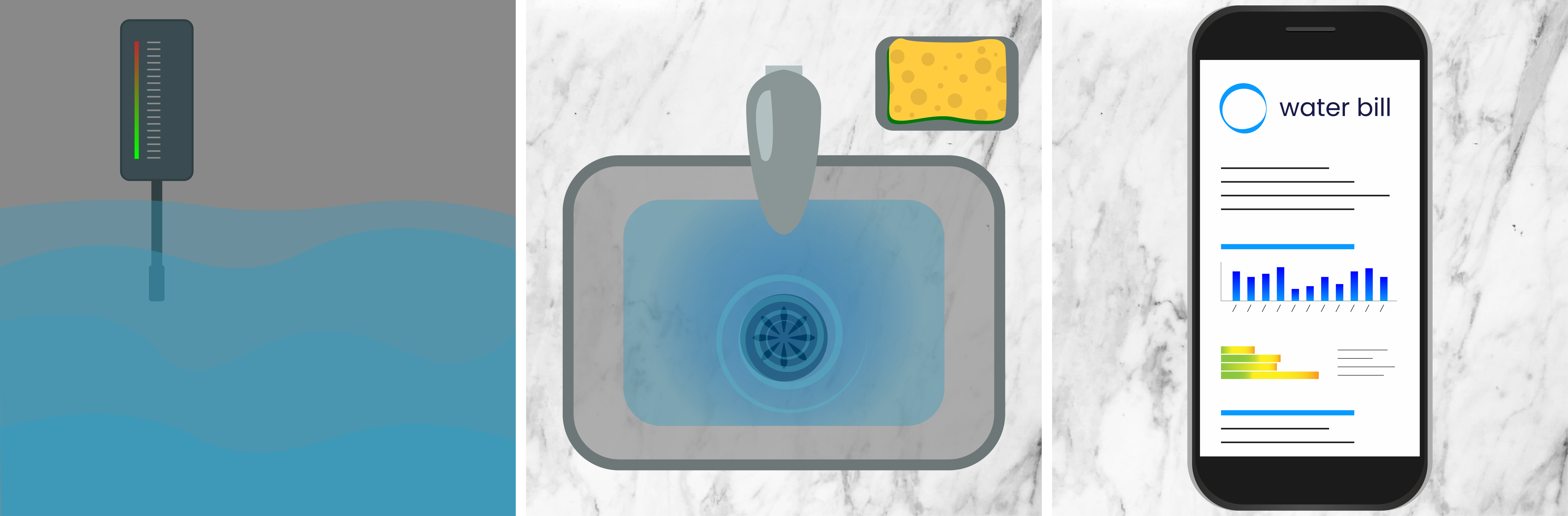}
        \caption{Scenario 3: Alex is keeping track of water consumption.} \label{fig4}
        \end{figure}
\end{itemize}

\subsubsection{Card-sorting task}
After the discussion, we wrapped up the factors by listing them on the room's big screen. 
The two moderators declared to the participants a list of privacy factors derived from their discussions and directly asked if anyone had any further potential factors to add to our list. Once all the suggestions had been documented, the two moderators asked the participants to rank the items that came out of their previous debates in order of importance using a method analogous to a card-sorting task. These factors were elaborated by the main and assistant facilitators, discussed and approved by all the participants. This procedure helped to clarify the priorities of the different focus groups.
%Then, we start with the task. Such a task follows the card-sorting task rules (e.g., rank the factors from the most to the less important in terms of privacy). We leave the users the possibility to explain what is the factor that they have chosen as most important and why.  

% We thank all the participants and we compensate them with a coupon for each that can be spent in the store or in the canteen of the university. Then, we verify the automatic transcriptions and save them on a secure hard disk.

\subsection{Thematic Analysis}
We followed the steps in~\cite{braun_2006_method} for the thematic analysis of the qualitative data obtained: 
\begin{enumerate}
    \item read the transcripts;
    \item underline any sentences or block of sentences related to privacy concerns;
    \item code the sentences from the previous step;
    \item merge the codes from the previous step in themes;
    \item review the themes and the related sentences to verify the homogeneity between them.
\end{enumerate}
All authors independently read the transcripts and coded the sentences. 
We then compared the codes and debated until a consensus was reached, then derived the themes and reported the results. 
% Thanks to the final list and the discussion with the participants about the factors, we were easily able to define the codes. 

\section{Results}\label{sec:results}
Our thematic analysis resulted in 9 themes that correspond to privacy factors: 
\textbf{transparency, control, trust, privacy of bystanders, risks, data minimization, confidentiality, privacy/security trade-off, potential misuse and unexpected purposes or consequences}. Table~\ref{tab5} is the summary of our themes and codes. They are related to users' privacy concerns, preferences and expectations when using IoT apps.
As also our results demonstrate, we reached data saturation within and across the three focus groups, because the observed issues and factors began to be repeated within and among the focus groups and further focus groups for data collection became thus redundant. 
In this section, we present the derived themes in the  subsections including the codes that we extracted from the transcripts and which are printed in bold. 
%The three transcripts from the focus groups count approximately 18000 words.
%We present in Appendix~\ref{appendix-1}, Table~\ref{tab3}, the examples provided by participants for the applications that indicate their understanding of the topic, elicited before starting the discussion. 
To characterize the quotes, we provide in Table~\ref{tab4} the demographics of the participants in our focus groups.

\subsection{Transparency} 
%+ Dashboard/overview (FG2)
%   + Ex ante: categories and (active) settings, automation rules, use and duration of collection (FG1, FG2, FG3
%   + ex-post (logs) (FG1, FG2)
%   + ex-post: data recipients, shared with whom (FG1, FG2, FG3)
In all three focus groups participants were concerned about the \textbf{Transparency} of the app both before and after the data disclosure and the execution of applications. Transparency before the automation 
%of 
by the app, in terms of \textbf{settings and rules}, such as active settings, automation rules, use and duration of the collection, P2 indicated: \textit{"what I meant with the person who programmed this app for you. I don't know what kind of information they get about how you use it. I don't know how that works"}. Transparency during the execution of the automation, a code that we derived is the \textbf{general overview} and presents the opportunity to get a summary of the Trigger-Action apps as P8 suggested: \textit{"I thought about the dashboard. I can visualize how the type of instruction that you give in the application."}. Transparency after the automation, the \textbf{data recipients} has been discussed, P2 briefly described it with \textit{"who gets the data"} and P7 voices concerns regarding the transparency of data recipients: \textit{"It's really weird if it would be invisible. Like what has it done? Where did the video go? Did it go to several places? Did it only upload you? You would feel weird if you didn't see". }

%The worries presented about individuals and organizations on the misusing of the information obtained, indeed P4 elaborated more by saying \textit{"if someone can respond to what's happening, then someone could know when you're home, when you're going home, when you're living, when you're on holidays, when you usually have food. If it's someone with bad intentions or if it's like a company knowing what you eat, what you order online, that kind of things, they could target you very, very well with advertisement, which wouldn't be beneficial for you eventually."}.

\subsection{Control} 
%Control before final action (upload, release), filter (FG1, FG2, FG3)
%  - no full automation,  “no sharing” as a default (FG2)
%  - notification instead of action (FG2)
%  - Consent
%  - Possible to object to data sharing for a time  period (FG1, FG2)
%  - Granular control (configuration) on triggers/actions (FG3)
%  - Granular control over the distribution of data (who receives what) (FG3)
The theme of \textbf{Control} was highlighted in all focus groups, especially in relation to scenario 2.  In particular, \textbf{control before the final action} was required, as P7 specified that \textit{"if I were Alex, I would definitely not like it
to upload immediately. It would be very scary if it just uploaded without you like pressing upload."}. 
The requirement that there is an option to filter information before it is uploaded was stated, e.g. by P2: \textit{There should be a filter}. 
%The previous code is well connected to the 
This is well related to the requirement of participants of \textbf{no full automation} that P8 exemplified by saying that \textit{"I think it's important to have this manual step or like, there must be some way to say This is not what I wanted."} and P7 replied: \textit{you still want to manually press the upload}. 
%These two codes were particularly visible during the discussion of the scenario 2 and both are connected to the importance of obtaining the user consent before automatically uploading any data. 
%Indeed the code \textbf{consent} was exemplified by P12 \textit{"this is horrible: record people without their consent."}. 
A suggestion came in focus group 2 from P8 and is related to \textbf{notification instead of action}, by observing that \textit{"the user will have still have some control, but this just helps to remind them that these things are coming out"}. 
The suggestion to \textbf{withhold sharing for a time period} for enhancing user control was brought by P11: \textit{Set a delay for the recording and uploading layer after record. You can set like uploaded 6 hours later, so if you don't like this video, you can just cancel to uploading it. }
%Also \textbf{stop sharing for a time period} is another code discussed that express the desire of setting a time frame when the applications can run, for example P2 \textit{"there should probably also be a time based kill switch. Turnoff. Stop recording after X time"}. 
Some participants suggested a \textbf{granular control in the configurations} and a \textbf{granular control in the data sharing}, especially with the ability to trigger the applications with a specific switch on a personal device indicated by P15: \textit{"if a sensor triggers something, I want to trigger something. If I have a smartwatch, I want a  button that says lower the temperature in my house to ten degrees"}. with the ability to trigger based on sensor data and to initiate the designed actions.

\subsection{Trust}
%  - device (FG1, FG2)
%  - controller, processors, recipients (FG1, FG3)
%  - Trustworthy safe service (FG2)
%  - Assurance guarantees (certification, trust/privacy seals and indicators) (FG1)
The theme of \textbf{Trust} was highlighted and covered by several participants in all focus groups, in the general discussion at the beginning and in relation to the different scenarios. P4 suggested that there could be a \textbf{device} designed with no privacy as a priority and it could be vulnerable: \textit{"a company might not be experts in data privacy IoT systems. They might come from more like hardware companies and a lot of devices cannot even be updated. So you have super old systems running that are very favorable, that are very vulnerable and all of your data are transiting through them."}. Not only on the device, indeed another code is the trust on \textbf{controller, processor, recipient} that was pointed out by P13 \textit{"it all depends of course the level of trust of the device, but at some point you need to trust someone. And if you say that you trust this company to install it."}. Also \textbf{trustworthy safe service}, as raised by P11 \textit{"a user has to consider this kind of service trustworthy and safe because they may use it for a longer term that they don't want to have a like information crisis"}. \textbf{Assurance guarantees}, including  certifications or trust indicators and seals, were  discussed in focus group 1. For instance, P4 indicated that \textit{"you need to use you need to have like trust. Also, that's more than trust. There are certifications."}

\subsection{Privacy of bystanders}
The theme \textbf{Privacy of bystanders } was highlighted in all three focus groups, mostly for scenario 2.
% 
 % +(lack of) transparency for bystanders (FG1, FG3)
 % + consequences of a user’s action on bystanders (FG3)
 %    + Severe consequences for minorities/vulnerable populations, minors (FG1, FG3)
 % + Need for protecting bystanders (FG1, FG3, FG2)
The ubiquitous interaction with the IoT devices and services open data collection risks also for non-device owners, as P2 raised: \textit{"your convenience is versus everybody else privacy is sort of feels like a pretty clear because as long as Alex is influencing his own life or her own life and not negatively or not impacting anybody else, I don't have a problem with it."}. Indeed, participants shared their concerns about the privacy of non-users as bystanders like people in a public place filmed through smart glasses without awareness as our scenario 2 described. The privacy concerns for the bystanders include the \textbf{transparency and consent for bystanders}, in particular as P4 from the first focus group described: \textit{"it's the lights should be on whenever it's filming, which is supposed to say to the people that the device is working as it's supposed to be"}. Furthermore, the \textbf{consequences of a user’s action on bystanders}, was discussed, e.g. P15 mentioned: \textit{"I think the big thing is that Alex presses record is automatically uploaded and with a bunch of people that doesn't want to be there."}. Another point highlighted was the severe \textbf{consequences for minorities or vulnerable populations,} when P13 described that \textit{"if Alex is recording something of he or she [and others]. What happens to them? But they might be able to record someone that might be in trouble. That say that they have some dissident from a foreign state that is running away from some authoritarian regime."}.
Also, the need for \textbf{protecting bystanders} was mentioned, for instance by P2: \textit{"you could detect people with some face recognition or something that detects that there's people in this in the clip and then don't upload it."}.

\subsection{Risks}
%- security risks (e.g. unwanted access, impersonation) (FG3, FG1)
%  - data leakage (FG1, FG3)
%   - Sensitive inference (FG1, FG3)
%  - safety of execution (FG2)
%  - Safety risks including health risks (FG1)
%  - hackers (FG2)
%  - accidental data sharing, misconfiguration (FG2)
%  - Reliability concerns for automated triggers (FG3)
%  - IFTTT, a single point collecting lots of data (FG3)
% - % sensitivity of data type
The theme \textbf{Risks} includes privacy, security and safety and potential threats, and emerged in all focus groups in the general discussion and for all scenarios. The \textbf{security risks} described the possibility of unwanted access and impersonation as P15 indicated \textit{"it has to be pretty secure because you can prank your neighbor a lot if you can connect to the neighbor devices."}. The \textbf{data leakage} and sensitive inference is another code that we derived, 
%indeed P6 suggested that \textit{"Alex buys such a device and has to accept that there's potential for such information to be disclosed. And you just have to decide whether you will accept that risk or not."}. 
indeed, P5 was concerned about: \textit{``the Chinese government backdoor device that you have bought, will know everything about it''}, and 
 P4 asked in the context of scenario 1a \textit{"do you want them to track how your heartbeat is going and to, like, adapt your insurance to it?"}. 
 \textbf{The safety of execution} was another point of discussion, in particular, P8 raised \textit{"these two services that are connected, maybe they will change over time and what is happening in one and might not connect the way you expect to the other end. Maybe a year later, what you planned for this connection to do is something completely different. And now it's doing something you didn't want it to do"}. 
 The risk of \textbf{hackers} is still part of the users' concerns, P8 indicated that \textit{"other people are trying to hack you or like if you connect certain services, maybe if you lose your password and someone else has it, then they can suddenly change things in your life."}. The actors in the system are also the end-users, and P7 described in part the theme \textbf{accidental data sharing} in relation to scenario 2 by suggesting that \textit{"there's been so many accidental uploads and stuff to the Internet so that we should learn that not to do this without care."}. The participants argued about the elaborated trigger that becomes part of the end-users' everyday life, the \textbf{reliability concerns for automated trigger} was suggested by P12: \textit{"so the computer has to make you quite complex decision. And my question is, Alex didn't like programming a lot of things, but maybe configuring this will take some effort."}. Also, the \textbf{sensitivity of data type} that was highlighted by P5 that described \textit{"there's just the confidential stuff on your phone."}. Also the issue of \textbf{surveillance} whether by commercial entities or the government, was also discussed and mostly during scenario number 3, when P3 indicated \textit{"if you live in a totalitarian state, maybe they could try to control how much water you use your system"}.

\subsection{Data Minimization}
%+ conditional/contextual access only, restricted duration (FG1, FG2)
%  + location restriction for use (FG2)
%  + Minimise data sharing (e.g. via aggregation, lower granular data) - less data means lower risk (FG1, FG2, FG3)
%  + PETS to minimize (e.g. encryption) (FG3)
Similarly, \textbf{Data Minimization} theme appeared as the result of discussions in all focus groups and for the different scenarios. Related to data minimization, \textbf{conditional and contextual access} through selective disclosure controls was discussed during the second focus group when participants indicated the preference of giving access to their data when needed without a continuous access, as P10 indicated \textit{"if you could have different settings, you can only look at my fridge when I'm using the app and not looking all the time."}. 
Relating to \textbf{minimize data to be shared}, strategies of minimizing data through data aggregation or generalisation (by releasing data with at a higher granularity) were discussed. For instance, from P4 we got the quote \textit{"trying to share the less information possible."} and
P1 suggests to rather share for instance less location information by sharing location data only with a higher level of granularity: \textit{``But the location could thus just be: Is your phone in the Wi-fi, or not?"}.
%P8 we got \textit{"sharing your information might result like any type of information can probably be used against the person themselves. Maybe I'm just not creative enough to come up with a scenario, maybe for every single data type, but there could be a concern of ours since there's the chance of some stuff."}. 

\subsection{Confidentiality}
%  - encryption of communication (FG1)
%  - surveillance (FG1, FG2, FG3)
%  - self-hosting (FG1)
The theme of \textbf{Confidentiality} appeared in all focus groups. Participants highlighted \textbf{encryption of communication}, as P4 observed \textit{"I think Alex would need to know that the exchanges are safe and that they're encrypted when they're getting through."} as well as P12 described: \textit{"if we're talking about the list being on the cloud, I mean, you can encrypt the data"}.  \textbf{Self-hosting} was discussed as a security measure for expert users in focus group 1, where P2 suggested that \textit{"if Alex is really serious about security, they could have their own server"}. Another code that emerged is \textbf{encryption of data at rest}  that protects stored data through encryption, P12: \textit{"being on Google Cloud, I mean, you can encrypt the data"}.

\subsection{Privacy/Security trade-off}
%tradeoff with usability/convenience (FG1, FG3)
%  - tradeoff with utility / more rules and less privacy (FG1)
In connection with \textbf{Privacy/Security trade-off} that emerged as the result of discussions in FG1 and FG3, the \textbf{usability and convenience} was discussed and participants are aware that the apps may not be completely secure, as P13 suggested \textit{"some kind of automated functions, I would think is very convenient, but maybe not so secure."}. Closely, the configurations process might be complex and with many rules wouldn't be simple to know the privacy implications, as in part P12 described \textit{"Tracking this huge system of automation that you have somehow built. Right? Some things are easy, but if you configure for every light in the home and then you also think, when I leave the home, I also want to do something else."}. Another code was \textbf{utility} in terms of the balance between the benefits and risks, then P9 suggested \textit{"I thought about the dashboard. I can visualize how the type of instruction that you give in the application. It could be helpful with somebody because I might want to read up and I want to most want to know that what type of instructions is it in the applications"}.

\subsection{Potential misuse and unexpected purposes or consequences}
%+ customer monitoring with unwanted consequences (FG2, FG3)
%+ Consequences: social stigma, shaming, unfair judgements (FG2)
%+ sensitivity due to unspecified context (FG2)
%+ unexpected data tracking/sharing/processing (e.g. advertisement, insurance companies, tracking, monetization) (FG1, FG2, FG3)
This theme, which emerged as the result of discussion in all focus groups, includes several codes, one of these is the \textbf{unexpected data tracking/sharing/processing} that arise when the end users feel the possibility of being tracked without awareness and consent. For instance, P7 mentioned that \textit{"maybe the device starts and it does something that they don't know and tracked data that they don't feel comfortable."}. Relating to scenario 1, we discussed \textbf{customer monitoring} and P8 indicated improper use of user data, in particular, that \textit{"if the store gets your information they will start like giving personal discounts to you based on what's in your fridge."} that is close to what P7 suggested \textit{"Maybe they can like try to influence you your decisions by like controlling your expenses."}. During the discussion of scenario 2, the theme of \textbf{sensitivity due to unspecified context} was raised by P7 who highlighted the difference between the insights of certain data types: \textit{"it's like it could be a video could contain anything. It's not just your vegetables, it's not just your water usage. It's going to be anything that you do in your life could be."}. In scenario 3, the \textbf{social stigma, shaming} was discussed and P7 indicated that inappropriate public exposition of data such as personal behaviors or habits may be uncomfortable for the users, indeed \textit{"if you're like someone who takes long baths, then you like to shower in hot water. Maybe you don't want to  see, but everyone knows and everyone can see."}
\newline
\newline
\indent During the final card-sorting task, participants revealed what privacy factors they thought are most important in the context of IoT trigger-action applications. In each of the three focus groups, we used different terms to determine the factors discussed, even if their meanings were close. It is relevant to mention that a factor can be a theme or a reformulation reported in this Section~\ref{sec:results}. Relating to the first focus group, there is no prevalence of particular factors, the participants selected the data controller, the security of the chain, and the sensitivity. These three factors, in order, reflect the meaning of the code \textbf{controller, process and recipient} under the theme \textbf{Trust}, the \textbf{encryption of communication} of the theme \textbf{Confidentiality} and the code \textbf{sensitivity of data type} under the theme of \textbf{Risks}. In the second focus group, there is a clear predominance of the code \textbf{control before the final action} under the theme of \textbf{Control}. In the third focus group, both the themes \textbf{Trust} and \textbf{Control} were indicated as the most important factor. 

\begin{table}[!h]
    \centering
    \caption{Themes (privacy factors) and codes}\label{tab5}
    \begin{tabular}{|c|l|} 
    \hline
    Themes & Codes \\
    \hline
    \textbf{Transparency} & \makecell[l]{Transparency of settings and automation rules \\ General overview  \\ Transparency of data recipients} \\ \hline
    \textbf{Control} & \makecell[l]{Control before the final action \\ No full automation \\ Notification instead of action \\ Withhold sharing for a time period \\ Granular control in the configuration \\ Granular control in the data sharing} \\ \hline
    \textbf{Trust} & \makecell[l]{Trust in the device \\ Trust in controller process recipient \\ Trustworthy safe service \\ Assurance guarantees }
    %(certification seals indicators)
    \\ \hline
    \textbf{Privacy of bystanders} & \makecell[l]{Transparency and consent for bystanders \\ Consequences of a user's action on bystanders \\ Consequences for vulnerable populations \\ Protecting bystanders} \\ \hline
    \textbf{Risks} & \makecell[l]{Security risks \\ Data leakage \\ Safety of execution \\ Hackers \\ Accidental data sharing \\Reliability concerns for automated trigger \\ Sensitivity of data type \\ Surveillance} \\ \hline
    \textbf{Data Minimization} & \makecell[l]{Conditional and contextual access \\ Minimize data to be shared} \\ \hline
    \textbf{Confidentiality} & \makecell[l]{Encryption of communication  \\ Self-hosting \\ Encryption of data at rest} \\ \hline
    \textbf{\makecell[c]{Privacy/Security \\ trade-off}} & \makecell[l]{Usability and convenience \\ Utility} \\ \hline
    \textbf{\makecell[c]{Potential misuse and \\ unexpected purposes \\ or consequences}} & \makecell[l]{Unexpected data tracking/sharing \\ Customer monitoring \\ Sensitivity due to unspecified context \\ Social stigma} \\ \hline
    \end{tabular}
\end{table}

\section{Discussion}\label{sec:discussion}
\subsection{Key privacy factors for IoT TAPs}
In this study, we contribute to illustrating the end-users' privacy concerns and preferences, and thus their privacy expectations regarding IoT Trigger-Action applications. Our analysis shows that
%, during the three focus groups, 
privacy factors in terms of privacy preferences and concerns that matter for users in the general IoT context, as derived earlier by related work and summarised at the end of section~\ref{Related}, and are reflected by our derived themes and codes. 

However, our findings go beyond the related work, as our scenario-based focus groups allowed us to elaborate and analyse in more depth what factors matter specifically for IoT TAPs and why they are important for users.

While previous work has revealed that transparency, trust and control are generally important privacy factors for IoT environments, our results show that in the IoT TAP context these factors especially matter in relation to the automation of data sharing with third parties (data recipients) and the automation of actions performed on these third parties. The automation of data sharing and of actions, which are specific to IoT TAPs scenarios, are resulting in new nuances of privacy preferences and concerns regarding the control and transparency of automation. Users are particularly interested to have control over the final action and to restrict or disallow full automation. 
They demand transparency of automation settings and rules and transparency of  data recipients. Moreover, they not only consider trust in the data controller as important but also emphasise the importance of trust in the automation process and in data recipients.

Additionally, unexpected and surprising automated actions resulting in  unforeseen sensitive data inferences and unexpected consequences  were also commonly shared concerns in all focus groups in relation to IoT TAPs, which could be caused by insufficiently controlled automation. Particularly unexpected automated actions that could unintentionally impact the privacy of bystanders and reveal sensitive information were mentioned as concerns in relation to scenario 3.

Hence, privacy controls including permission systems have to address these key privacy factors that matter for users when using IoT TAPs.

%standard themes in the general IoT context, such as confidentiality, risks (e.g., leakage of sensitive data), user control, privacy of bystanders, and transparency and more dedicated to the Trigger-Action apps, such as control before the final action and no full automation, unexpected consequences, trust in controller, process, recipient, and transparency of data recipients were underlined.
%such as transparency, control, risks, confidentiality, trust, privacy of the by-standers, data minimization, privacy/security trade off, and potential misuse and unexpected purposes or consequences. 
In this Section, we further discuss and examine the key findings 
%to characterize our RQ.
relating to our RQ.

%bystanders.
\paragraph{Privacy of bystanders}
Consistent with previous works regarding the context of smart home, a lot of attention has been received by bystanders and guests, also known as secondary users, incidental users~\cite{marky2020visitors,cobb2021incidental,yao_2019_bystanders} and to concerns related to AR glasses~\cite{lebeck_2018_ar}. The smart home scenario is one of the possible environments where end-users can set, implement and utilize Trigger-Action applications. In our focus groups, and in particular, in the scenarios related to smart city and wearables, participants were concerned that the actions taken by TAPs may impact other individuals who are not directly involved in the TAPs. Such attention on by-standers emerged at most during the discussion of scenario number 2 and number 3, where the automation of uploading a video recorded emphasized the privacy concerns such as the lack of transparency, and the consequences for the non-device owners. 
%When a collection of information about people who the consent may not have been given to the collection is inappropriately made, there is the need of protection for those who are not actively using the service or device.

%control & trust
\paragraph{Trust and control over automation}
Users have a sense of privacy protection when they trust and feel in control of their data. Initial observations suggest that there may be a link between the level of trust and perceived privacy risks to providing personal information~\cite{kim_2019_trust}. Indeed, control before the final action is a user's need, specifically for IoT apps. 
For small companies it may be expensive to update the IoT devices to fix the vulnerabilities~\cite{meredydd_2016_iot}, on the other hand, the users are conscious about the exposition of their private data that flow between devices, indeed the trust in the device contributes to their worries. Obtaining certifications can enhance reliance on a data controller to understand the decision-making process, data recipients, and data flow within TAPs, which can be complex due to the presence of multiple services and devices~\cite{kulyk_2020_trust}.
To gain more control, our results show that users would have the option to stop sharing for a time period. In the TAPs context, there is the opportunity to define spatial and/or temporal exceptions and they may include the stop sharing in a set of circumstances~\cite{desolda_2017_filter}. Another relevant concern was the need to have control before the final actions, in terms of review and approval, before any action is executed since the interconnection of two services/devices in TAPs may introduce a complex data flow between them.  Similarly, the granular control of configurations and data sharing demands the ability to get the opt-in to data sharing on a granular level and in sensitive situations to receive a notification instead of a full automation to review the proposed action and make an informed decision. %Likewise, the consent is a priority in terms of providing clear information about data collection and usage practices, with mechanisms that permit the users to give or withdraw consent in any moment.

%risks
\paragraph{Sensitive inferences}
Users don't have a full awareness of machine learning and data inference techniques for profiling~\cite{chow_2017_iot}. Similarly, our findings show that the risk of data leakage is connected to the risk of sensitive inference. Moreover, in our focus groups, participants suggested the potential derivation of private information that can reveal delicate data thanks to a statistical inference of the data shared. One of the worst privacy panic scenarios involves the security risks or hackers~\cite{angulo_2015_panic}, which include unauthorized access to their personal information that could lead to harm to individuals, including identity theft, or execute unwanted actions. On the other hand, relating to safety, when the end-users involve in the execution of physical actions, such as unlocking the house's doors when the owner is not at home or simply turning on lights, the concerns are more about the potential of physical damage resulting from errors in execution. One of the most discussed concerns, in particular during scenario number 3, is related to surveillance, when the participants were worried about the public controller, such as a government, that can derive sensitive information, similarly to~\cite{naeini_2023_city} in the smart city context. %The accidental data sharing is a consequences of unwanted data sharing permissions given as well as unsecure storage or transmission that could expose end-users data online. The sensitivity of data type is another relevant aspect that includes different users' perceptions of the data shared, such as biometric or location data that may be more sensitive and require more attention to mitigate any privacy risks. 

%transparency.
\paragraph{Transparency of automation}
Users tend to feel more confident in their activities with IoT technologies when they have access to easily understood information about how their data is being collected and used. 
Transparency is one of the most important factors for IoT users, while for data protection and IT professionals, the most important constructs are anonymity, simplicity, explicit consent, and GDPR compliance~\cite{cormican_2022_iot}. 
In TAPs, the end-users could share their data with more than one data recipient if the Trigger-Action application connects two different entities. The data recipients could include third-party developers, advertisers, or other individuals or organizations that may have access to the data collected through the application. Other suggestions from the participants are related to the general overview of the applications installed, in a way that the end-users can access the information of the data shared in an easily accessible and user-friendly form. %Closely, the transparency in settings and rules to get a clear view of the behavior of the apps, such as when a trigger and actions execute, and for how long data will be stored.

\paragraph{Unexpected consequences}
Additionally, unexpected consequences and the potential for misuse  may undermine users' adoption of these IoT platforms, and ultimately impact their overall privacy perceptions even more since two different entities should work as expected. In Trigger-Action applications, the presence of multiple interconnected services increases the likelihood of unintended exploitation of users' personal data, leading to issues such as monitoring and sharing without their consent. Moreover, users may experience social shaming if their personal habits are made public. The exposition of user data, for example of water consumption that is our scenario number 3, is a potential social stigma. 

\subsection{Towards usable privacy permission settings}
The derived privacy factors will
%as we mentioned in Section~\ref{sec:introduction}, 
serve as an input for a follow-up quantitative study for finding typical profiles of privacy permission settings that can be offered to users and may be used to support usable privacy permission management for IoT TAPs. 

While users should start by default with the most privacy-friendly permission settings in compliance with the Data Protection by Design and Default principle of Art. 25 of the GDPR, providing them with predefined and selectable profiles of permission settings including profiles that are also likely matching their preferences, will allow them to more easily change and adapt their settings.

Machine learning (ML) techniques on the user’s (mobile) device (i.e., under their control) could be used to evaluate the users’ decisions for predicting and suggesting privacy profiles for IoT TAPs that are best fitting the user's behavior regarding allowing or denying privacy permission requests,
similarly to personalized privacy assistants presented for instance by ~\cite{liu_follow_2016,bahirat_data-driven_2018,smullen_best_2020} that were developed for mobile and IoT environments.
Such an ML-supported approach can be combined with protocols enabling ``on-the-fly'' permission management,
similar to~\cite{angulo2012towards}, which can support users to easily adapt their permission settings ``on-the-fly'' based on the users’ behavior and predicted needs, and relating to the user's current context.

\section{Limitations}

Our study has limitations common to many user studies and to user studies in the area of privacy.
% \subsection{Recruitment bias}
% Some may have guessed it was about privacy although we carefully avoided to prime them.
\subsection{Diversity in the focus groups}
First, we acknowledge an imperfect diversity among the participants of our focus groups (see Table~\ref{tab4}).
More precisely, 12 out of 15 participants were less than 34 years old, and the three participants over 45 all belonged to the third focus group.
Most of our participants were also educated at a university level (14 out of 15 possess a Bachelor's degree or more), which is not representative of the country in which the study has been conducted, nor of the EU and North America as a matter of fact.
The relative youth and education of our participants are likely due to the recruitment, partially affected via university networks.
However, we contend that these two factors are not critical in that
\begin{inparaenum}
    \item participants were nonetheless spanning over various age groups, and
    \item educated people are likely to be IT users, and can therefore reflect privacy concerns and preferences of a larger portion of population.
\end{inparaenum}

% It is also possible to observe a small imbalance in gender, and especially in the second focus group were all participants were female.
% We did not observe specific differences between the groups that could be explained through gender.

\subsection{Lack of experience in IoT apps from participants}
Second, we note that none of the participants actually used an IoT app before.
Participants may have been asked about situations in which they are not typically put, therefore influencing their decisions by asking them to imagine how a fictional character (the Alex persona) would react.
In spite of this limitation, the focus groups allowed us to explore situations that \textit{may} happen.
For a novel field such as IoT applications, it may be necessary to elicit concerns and preferences for prospective situations in order to shape their design and inform regulators about future practices.

\section{Conclusion}\label{sec:conclusion}
Understanding privacy preferences and concerns that users have regarding IoT TAPs is important for designing usable privacy controls for this expanding technology. 
Our qualitative research composed of scenario-based focus group discussions revealed that privacy factors that matter for users of general IoT environments also matter for them for IoT TAPs. However, our explorative work also showed that users also have specific preferences and concerns regarding the automation aspect of IoT TAPs. They demand control and transparency of the automation process, regard trust in the automation process and data recipients as important, and are concerned about unforeseen privacy risks and consequences caused by the automation for themselves and for bystanders.

To gain a more comprehensive understanding of users' privacy preferences and concerns in the context of IoT TAPs, we intend to conduct a survey as a future study. The triangulation of data will allow us to validate and augment our qualitative research findings, providing a more robust basis for designing effective usable privacy permission management.

\section*{Acknowledgements}
This work was partially supported by the Wallenberg AI, Autonomous Systems and
Software Program (WASP) funded by the Knut and Alice Wallenberg Foundation.
%This work was funded by the CyberSecIT project as part of the Wallenberg AI, Autonomous Systems and Software Program (WASP) of the Knut and Alice Wallenberg Foundation and was 
It was also partially supported by the TRUEdig project funded by the Swedish Knowledge Foundation. We thank all focus group participants for their valuable contributions.

\bibliographystyle{IEEEtran} 
\bibliography{Mybib2.bib}
\appendix
\label{appendix-1}
\section{Details of study design: questions asked}
In this section, we present the exact questions we asked in different parts of the study including the demographic questions, questions asked at the end of the introductory session (part b, see Section~\ref{FGsteps}) to assure participants' understanding of persona, questions asked for the general discussion (part c), and scenario-related questions (part d). 

\textbf{Part a: demographic questions:} 1) What is your age? (Options: 18-24, 25-34, 35-44, 45-54, 55+, Prefer not to say). 2) What’s your highest level of education? (Options: PhD, Master, Bachelor, High School, Have not completed High School, Prefer not to say). 3) What’s your gender? (Options: Female, Male, Non-binary, Prefer not to say). 4) How would you judge your IT (Information Technology) knowledge? (Options: Excellent, Very Good, Good, Fair, Poor). \\
\textbf{Part b: questions to assure understanding of persona/context:} 1) Is everything clear concerning Alex? 2) Do you have any questions? 3) Can you imagine another situation where Alex would connect two entities? Would you describe that situation?\\
\textbf{Part c: questions concerning the general discussion on TAPs:} Now you are Alex, and you would use this kind of Trigger-Action application. How does Alex proceed? What Alex should consider? (\textit{Follow-up questions ordered based on their priority:}) 1) If you were Alex, what would be your concerns regarding the usage of the TAP? 2) What decisions Alex should make? What factors Alex should consider when using the TAP? 3) If you were Alex, what problems would you have when using a TAP? 4) What would be the benefits of using the TAP for Alex? 
5) If you were Alex, what functionality or features would you expect from a TAP? 6) What are the first words that come into your mind now that you have heard about such TAP applets? 7) Does the level of experience of Alex matter? \\
\textbf{Part d: questions concerning scenarios:} (After each scenario we asked) 1) If you were Alex, what would you think about the scenario? 2) If you were Alex, what would you like and do not like about this scenario? 3) If you were Alex, would you feel comfortable? 4) If you were Alex, would you set up the same rule? Why? 5) Imagine that you have set such a rule. If you were Alex, would you like to be informed about the data flow? In which way? (\textit{Follow-up questions for each scenario ordered based on their priority:}) 1) What type of personal information do you think is being collected in this scenario? 
2) Who do you think is collecting your personal information? Where do you think your personal information will end up?    
3) In what ways do you think your personal information will be used? Do you see any potential risks?    
4) Is privacy involved here?    
5) What factors would you consider as privacy-related here? 
  
\section{TAPs examples showing participants' understanding of the context}
\label{appendix-2}

%\begin{table}
%\caption{Participants Trigger-Action Applications examples.}\label{tab3}
%\begin{tabular}{|l|l|}
%\hline
%Participants & App description\\
%\hline
%P2 & if you go back home, they turn back on [lights].\\ \hline
%P3 &  \makecell[l]{if you go running and then you're coming home, \\ then the sound starts on.}\\ \hline
%P4 & \makecell[l]{you could have a wearable that detects a fall \\ and call the ambulance.}\\ \hline
%P5 & \makecell[l]{when I enter my home door, order me food \\ and I have like a plan  of what I want to eat every day.}\\ \hline
%P7 & \makecell[l]{when your pulse is going down and you fall \\ asleep and then it turns off your phone to sleep mode.}\\ \hline
%P8 & \makecell[l]{you are arriving at the bus stop. \\ I will purchase a ticket for you so you can go.}\\ \hline
%P11 & \makecell[l]{She or he could decide when he comes \\ back home and the room could be heated.}\\ \hline
%P13 & \makecell[l]{alarm in general in my house, fire, using electricity, \\ freezer is too hot [I would receive a notification]} \\ \hline
%P14 & \makecell[l]{the trigger could be the level of the pellets in the house, \\ then the action could be a notification on the smartphone.} \\ \hline
%\hline
%\end{tabular}
%\end{table}





% \end{thebibliography}

\end{document}